\documentclass[preprint, superscriptaddress, showkeys, floatfix, nobalancelastpage]{revtex4}
\usepackage{amsfonts}
\usepackage{amssymb}
\usepackage[dvips]{graphicx}
\usepackage{color}
\usepackage{amsmath}
\usepackage[bookmarksnumbered, bookmarks, breaklinks, linktocpage]{hyperref}

\begin{document}

\title{Ground state properties of two spin models with exactly known ground
states on the square lattice}

\author{A.~Sindermann}

\affiliation{Theoretische Physik, Universit\"at zu K\"oln, Z\"ulpicher
 Str.77, 50937 K\"oln, Germany}

\author{U.~L\"ow}
\affiliation{Theoretische Physik, Universit\"at zu K\"oln, Z\"ulpicher
 Str.77, 50937 K\"oln, Germany}

\author{J.~Zittartz }
\affiliation{Theoretische Physik, Universit\"at zu K\"oln, Z\"ulpicher
 Str.77, 50937 K\"oln, Germany}

\date{\today}

\begin{abstract}
We introduce a new two-dimensional model
with diagonal four spin exchange and an exactly known
ground-state. Using variational
ans\"atze and exact diagonalisation we calculate upper and lower bounds
for the critical coupling of the model.
Both for this model and for the Shastry--Sutherland model
we study periodic systems up to system size $6\times6$.
\end{abstract}


\maketitle
\pagebreak

\section{Introduction}
\label{sec:Intro}

In this paper we study two frustrated two-dimensional quantum spin
models with exactly known singlet dimer ground states.

\begin{figure}
\begin{center}
\includegraphics[width=8cm]{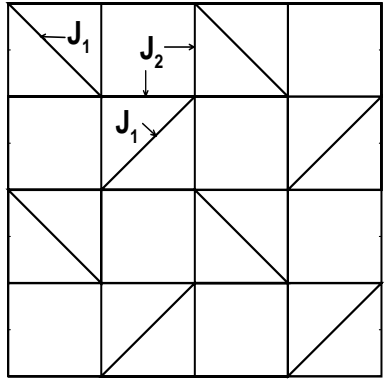}
\end{center}
\caption{Shastry--Sutherland model with $J_2$ along diagonal bonds
and $J_1$ along vertical and horizontal bonds. The dimer ground state is
build up by dimers i.e. $\frac{1}{\sqrt{2}}( \uparrow \downarrow -\downarrow \uparrow )$ pairs
along the diagonal bonds.} \label{fig1}
\end{figure}

First we consider the Shastry--Sutherland model
\cite{ShSu}, a two dimensional Heisenberg model with an additional
frustrating coupling of strength $J_1$ on every second diagonal
bond (see Fig.\ref{fig1}) with Hamiltonian

\begin{eqnarray}
  \label{eq:ham_ShSu}
  H&=&  J_2 \sum_{<i,j>}{\vec S_{i} \vec S_{j}}
      + J_1 \sum_{<i,j> diagonal}{\vec S_{i} \vec S_{j}}.
\end{eqnarray}
\noindent
Though the model shows similar properties also for higher spins,
here we restrict ourselves to the case of spin $\frac{1}{2}$, 
and the $\vec S_i$ denote spin-1/2 spin operators located at site $i$. The model
Eq.\ref{eq:ham_ShSu}  has attracted attention recently, when
it was suggested, that the magnetic properties of the substance
SrCu$_2$(BO$_3$)$_2$  \cite{Kage} are very well described by its
frustrated spin-spin interactions.

\begin{figure}
\begin{center}
\includegraphics[width=10cm]{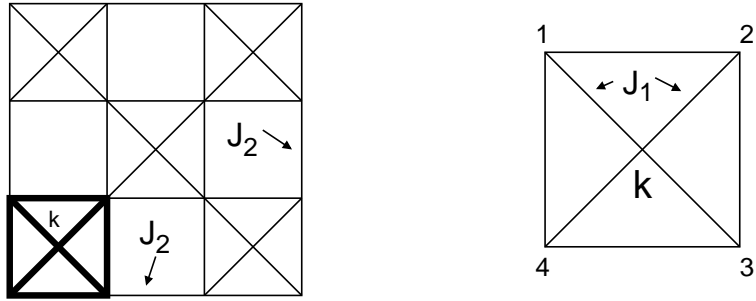}
\end{center}
\caption{Plaquette model with four spin interaction.} \label{fig2}
\end{figure}

The second model we investigate incorporates a diagonal ring
exchange term on every second plaquette (see Fig.\ref{fig2}). It
is given by the Hamiltonian

\begin{eqnarray}
  \label{eq:ham_Plaq}
  H&=&  J_2 \sum_{<i,j>}{\vec S_{i} \vec S_{j}}
      + J_1 \sum_{plaquette\ k} ({\frac{3}{4}+\vec S_{k_1} \vec S_{k_3}})( {\frac{3}{4}+\vec S_{k_2} \vec S_{k_4}})
\end{eqnarray}
\noindent
where again the $\vec S_i$  denote spin-1/2 spin operators and the
second sum runs over every second plaquette $k$ as depicted in
Fig.\ref{fig2}. The four-spin interaction is normalized in such a
manner that the dimer ground states, which play a major role in
what follows, have zero energy. The extra four spin interaction of
Eq.\ref{eq:ham_Plaq} can be viewed as a ``truncated'' version of
the ring exchange which has been  discussed in the context
of the two-dimensional mother substances of High $T_c$  \cite{Coldea}
and which have been found to play an essential role in spin ladder
substances like Sr$_{14}$Cu$_{24}$O$_{41}$ \cite{Schmi}.

Both models display rich phase diagrams and both have exactly
known singlet dimer ground states. The latter property is
otherwise rarely encountered  when dealing with two-dimensional
quantum spin systems: The ground state of the Shastry--Sutherland
model is non degenerate and consists of (singlet) dimers located
along the diagonal bonds, for the plaquette model the
corresponding ground state is two-fold degenerate, and each ground
state corresponds to a covering of the two types of diagonals with
dimers.

To facilitate the following discussion we define 
for both models the variable
$x:=J_2/J_1$, which can be viewed as an inverse frustration. The
models undergo a zero temperature first order phase transition to
the exactly known singlet dimer state at a critical inverse
frustration $x_c$. The numerical value of $x_c$ is 
different for the two models.

The aim of this paper is to introduce the basic properties of the
plaquette model and to determine the value of $x_c$ with best
possible accuracy for both the Shastry--Sutherland and the
plaquette model. Hereby we want to demonstrate, that also for
two-dimensional frustrated models which are hard to tackle
numerically, by use of the Lanczos technique on finite
clusters  not only reliable but also precise results for the
infinite system can be obtained. We also hope, that our results
are of use to gauge methods using uncontrolled approximations,
whose errors are usually hard to assess.

The paper is organized as follows. Sect.\ref{sec:plaq} is devoted to a study of
the plaquette model using variational methods and finite cluster
analysis. In particular two types of variational ans\"atze are used
to give upper and lower bounds for the model´s critical coupling
$x_c$. In Sect.\ref{sec:peri.1} we present a study of the Shastry--Sutherland
model of unprecedented precision using the Lanczos algorithm for
system sizes up to $6\times6$ with periodic boundary conditions.
In Sect.\ref{sec:peri.2} we present a similar analysis for the
plaquette model on periodic systems and give a comparison of the
two models. In Sect.\ref{sec:summ} we summarize our results and give an
outlook to future possibilities. 

\begin{figure}
\begin{center}
\includegraphics[width=8cm]{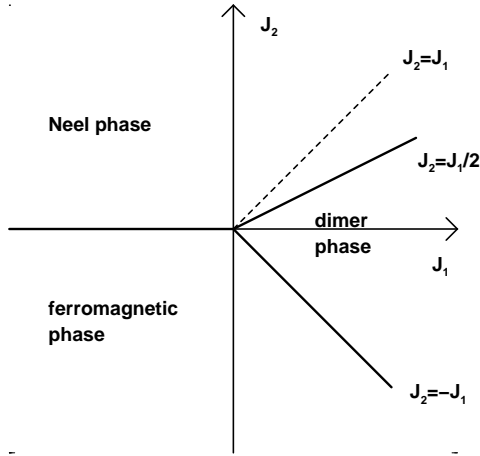}
\end{center}
\caption{Ground state phase diagram of the plaquette model.
The phase boundaries between the dimer 
and the N\'eel phase is drawn at $J_2=J_1/2$ which corresponds
to a system with $N=4$ spins.} 
\label{fig3}
\end{figure}

\begin{figure}
\begin{center}
\includegraphics[width=10cm]{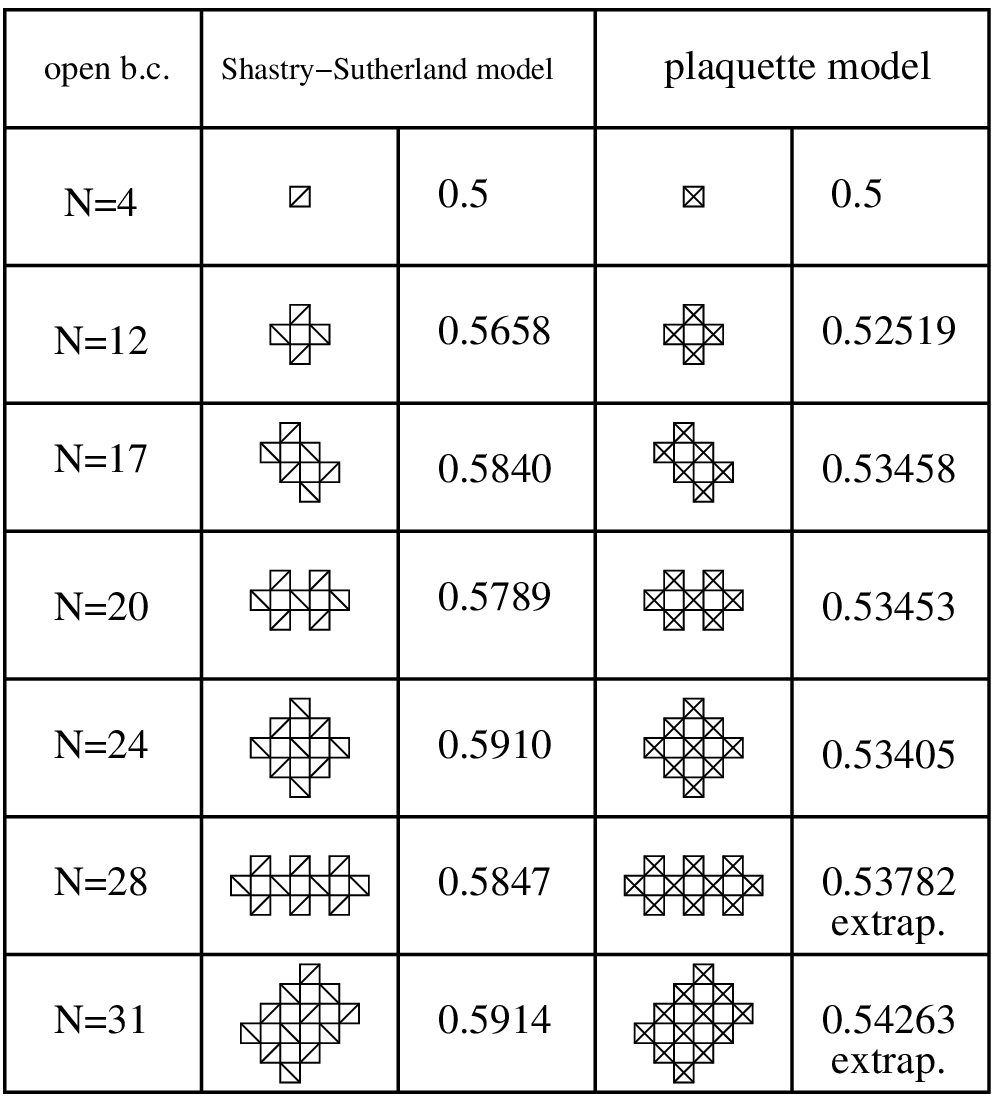}
\end{center}
\caption{$x_c$ calculated on finite clusters with open boundary
conditions for the Shastry--Sutherland model Eq.\ref{eq:ham_ShSu}
and the plaquette model Eq.\ref{eq:ham_Plaq}. The values for the
Shasty--Sutherland model are taken from Ref.\cite{us} and are
listed here for comparison.} \label{fig4}
\end{figure}

\begin{figure}
\begin{center}
\includegraphics[width=10cm]{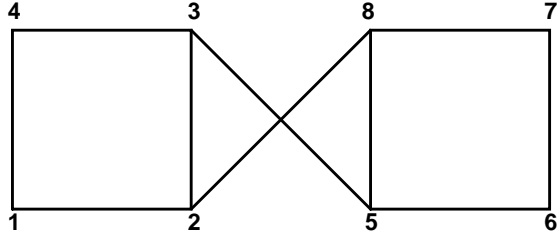}
\end{center}
\caption{A variational configuration 
of the "good type", consisting of 
two empty plaquettes  with four spins
and a four spin exchange between the plaquettes,
gives an upper bound  of 
$\frac{31}{48}$. For
a corresponding configuration of the Shastry--Sutherland model \cite{us} 
one obtains an upper bound of $\frac{3}{4}$, wheras a  configuration with 
the diagonal coupling $J_1$ on the plaquettes
("bad type") gives only $\frac{7}{8} $ as upper bound.}
\label{fig5}
\end{figure}

\begin{figure}
\begin{center}
\includegraphics[width=10cm]{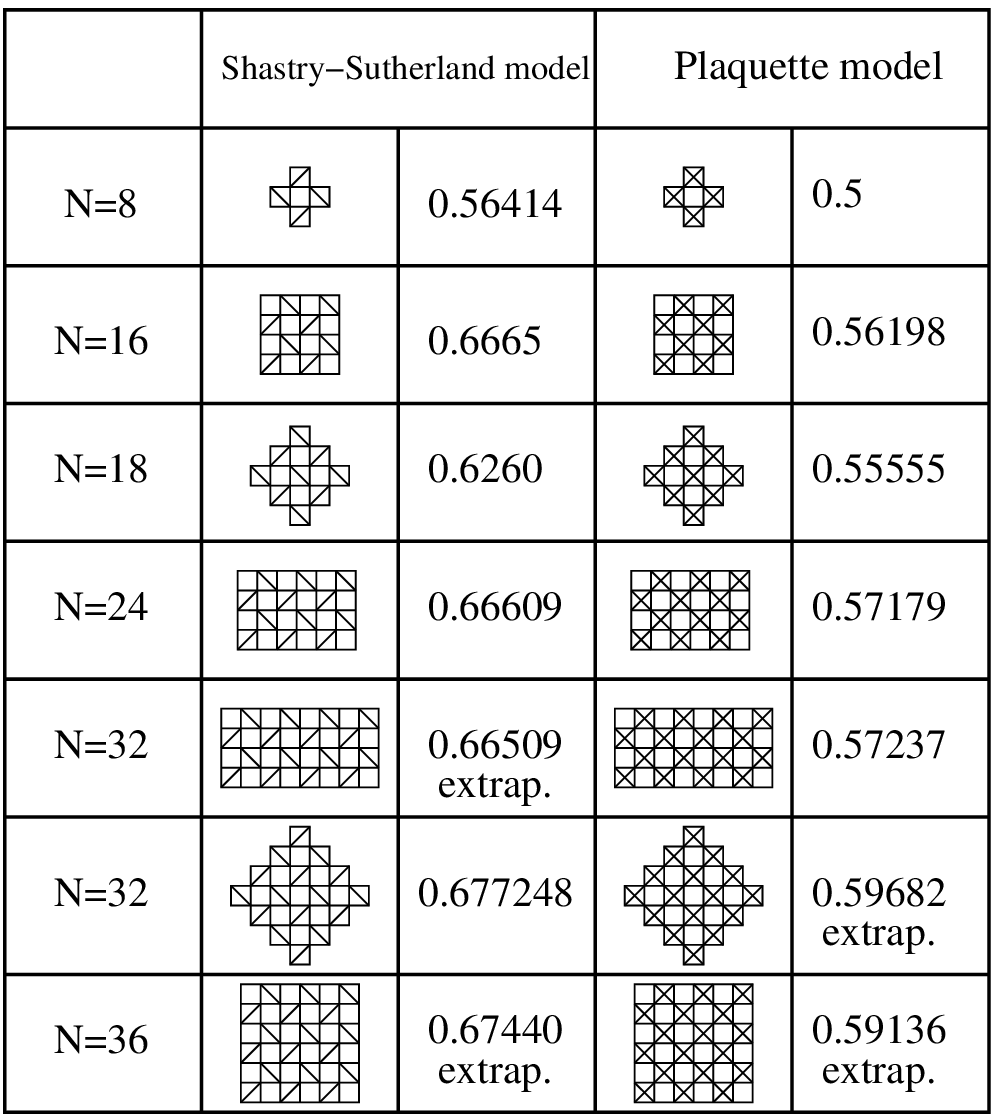}
\end{center}
\caption{$x_c$ calculated on finite clusters with periodic
boundary conditions for the Shastry--Sutherland model
Eq.\ref{eq:ham_ShSu} and the plaquette model Eq.\ref{eq:ham_Plaq}.}
\label{fig6}
\end{figure}

\begin{figure}
\begin{center}
\includegraphics[width=8cm]{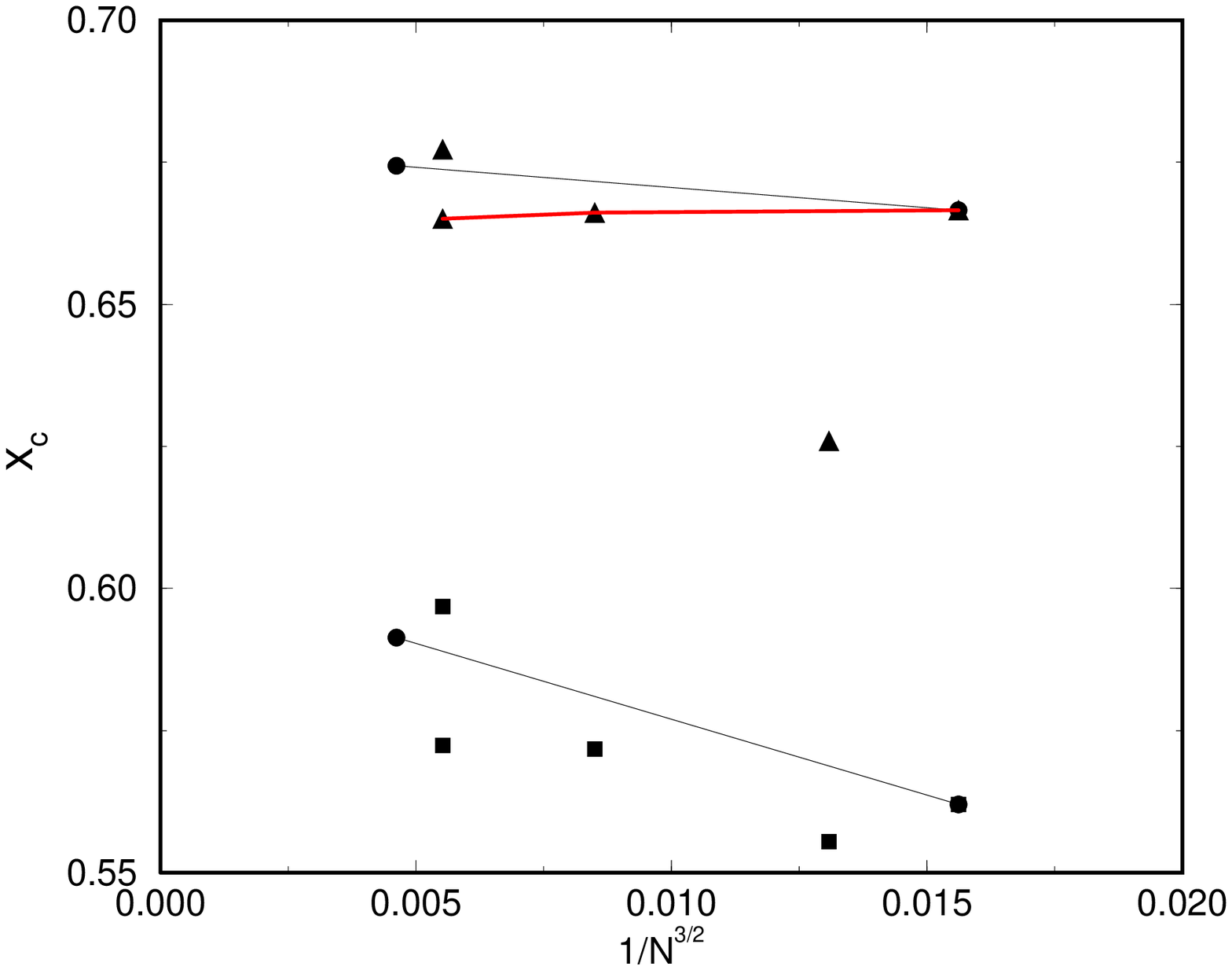}
\end{center}
\caption{$x_c$ plotted versus $1/N^{3/2}$ for the Shastry--Sutherland model
Eq.\ref{eq:ham_ShSu} (triangles and circles) and for the plaquette model
Eq.\ref{eq:ham_Plaq} (squares and circles). The lines connect systems
with the same shape} \label{fig7}
\end{figure}


\section{The Plaquette model}
\label{sec:plaq}

\subsection{The ground state phase diagram}
\label{sec:plaq.1}
 Some first insight in the model can be gained by
considering the smallest possible cluster, which consists of a
plaquette of four spins only. This system has two singlet
eigenstates with spin $S=0$, three triplets with $S=1$, and one
``ferromagnetic'' state with $S=2$. The ground state diagram can be
easily drawn for this system (see Fig.\ref{fig3}). Moving from the singlet
dimer phase in anticlockwise direction at
$x=\frac{J_2}{J_1}=\frac{1}{2}$ one enters a second singlet phase
usually referred to as antiferromagnetic phase or N\'eel phase and at $J_2=0,
J_1<0 $ starts the
ferromagnetic phase, which at $x=-1$ meets again the singlet dimer
phase. For $N=4$ the singlet dimer phase is degenerate with two $S=1$
phases.
The third $S=1$ state is a ground state only for $J_2=0$ and $J_1<0$,
where it is degenerate with the N\'eel and the ferromagnetic
state.

It is still an open question to what extent a third phase phase
between the dimer and the N\'eel phase exists for the infinite
system. This problem is similar to the still unsolved issue
in the Shastry--Sutherland model, for which various scenarios of
intermittent phases between the N\'eel phase and the singlet dimer
phase have been discussed.

The phase boundaries between the N\'eel phase and the
ferromagnetic phase and between the ferromagnetic and the dimer
phase are exact, i.e. they coincide with the boundaries of the
infinite system.
 However the value $x^{(N=4)}_c=0.5$ of the phase
boundary between the dimer and its adjacent phase is not correct
for the infinite lattice. To determine the value of $x_c$ also for
the infinite lattice, we calculated both lower and upper bounds
and also study finite clusters with periodic boundary
conditions.

\subsection{Lower and upper bounds for the Plaquette model}
\label{sec:plaq.2}
 Using variational ans\"atze in Ref.\cite{us} 
lower and upper bounds for the critical values $x_c$
were discussed for the Shastry--Sutherland model. Here we perform a
similar analysis for the plaquette model. As was pointed out 
by P.W.~Anderson \cite{Anderson} the ground state energy of open
clusters $H_{i}$ with no overlapping bonds, which cover the whole
plane, represent a lower bound for the ground state energy of the
infinite system. Similar to the Shastry--Sutherland system we
can infer thus from the 4 spin plaquette 
that $x^{(4)}_c=0.5$ is a lower bound for $x_c$ of the infinite
system. An analysis of systems of larger size (see Fig.\ref{fig4}) 
gives as best lower bound $x_c>0.54263$, which is the
critical coupling of an open system of size $N=31$. Unfortunately
the unsystematic variation of the bound with the system size does
not allow  a meaningful extrapolation.

To obtain an upper bound on the critical coupling $x_c$ the
Hamiltonian is split into clusters without common spins and into
external bonds connecting the clusters
\begin{equation}
  \label{eq:upper}
 H=\sum_{cluster} H_{cluster}^N +H_{bond}.
\end{equation}
This is graphically depicted in Fig.\ref{fig5} for clusters with
four spins. As variational state we use a product state
$\prod_{cluster} |\Psi_{cl}\rangle$, where $|\Psi_{cl}\rangle$  is
the ground state of the cluster with Hamiltonian $H_{cluster}^N$.
In contrast to the Shastry--Sutherland model however, where the
expectation value of the external bonds vanishes for the ground
state, which has spin $S=0$, in case of the plaquette model the
expectation value
\begin{equation}
\langle \prod_{cluster} \Psi_{cl}|H_{bond}| \prod_{cluster} \Psi_{cl}\rangle
\end{equation}
has to be explicitly evaluated. Again a first bound can be found
by considering clusters of size $N=4$. Best results are obtained
here by using a plaquette with no  four spin interaction as basic
cluster (as shown in Fig.\ref{fig5}). One thus obtains
$x_c<\frac{31}{48}\approx 0.6458$.
Again the numerical value can be improved here by considering larger clusters.
Since the situation is more complicated in the case of the 
plaquette model, because one needs 
to consider also the bonds, we restrict ourselves here
to a square lattice of size $N=16$, from which a numerical
diagonalisation gives $x_c< 0.627(1)$.

To conclude, we find $0.54263<x_c<0.627(1)$ as exact bounds on the
critical coupling of the plaquette model.

\section{The Shastry--Sutherland and the plaquette model on systems with periodic boundary
conditions  }
 \label{sec:peri}

\subsection{The Shastry--Sutherland model}
\label{sec:peri.1}

We do not intend to give a detailed description of the
Shastry--Sutherland model here since over the years it has been
vastly discussed (see e.g. \cite{ShSu,Kage} ). In this paper we summarise
only briefly the results on $x_c$, which are most relevant for our
work based on Lanczos technique \cite{Lan}.

Using variational approaches \cite{us} it can be inferred from
clusters of 31 and 32 sites, that $0.5914<x_c<0.6955$. Thus by
calculating variational bounds the margin left open for $x_c$ is
approximately 0.1. 

In this paper we discuss finite systems with periodic boundary
conditions, which usually are closer to the infinite size system 
than systems with open boundaries. Also they allow to study larger
systems, because translational invariance can be used to further
reduce the size of the subspaces
(for a list of the symmetries used for the different systems
 see table \ref{app_tab}).

 Here we concentrate in
particular on a precise determination of the critical coupling of
a $6\times6$ system. To successfully study this problem we had to
deal with subspaces of dimension $5 \cdot 10^{8}$. We accomplished
this by using parallel OpenMP programming techniques.
We also investigated a
$N=32$ square system and periodically closed $4 \times n$ systems
with up to $n=8$. An overview of all the clusters with
periodic boundaries is given in Fig.\ref{fig6}. Our results for
the $6\times6$ system are shown in table \ref{s36p_tab} and for the
$N=32$ square systems in table \ref{s32p_2_table}.

As a technical point we mention that except for the trivial case
of $N=4$, there is no argument why the ground state energy for
$x>x_c$ should be linear in $x$. To obtain optimal results 
we have therefore chosen values
of $x$ as close as possible to $x_c$ (see table \ref{s32p_2_table} and \ref{s36p_tab}).
This is most important for the $N=36$ system, for which
calculating extra points is time consuming. We thus find as result
for $N=36$ a critical value of $x_c=0.674408$, 
for the $32$ square we obtain $x_c=0.67724$ and for the $4\times
8$ stripe $x_c=0.66509$. More values for smaller systems can be
found in Fig.\ref{fig6}.

When studying two-dimensional clusters with periodic boundary
conditions it is obvious, that the finite size effects depend not
only on the number of spins in the system, but also on the
geometry of the clusters.  Here we have two types of systems, one
consists of the $N=4\times n$ stripes
with $N=16,24,32$ spins and
square systems with $N=16,36$. As was already pointed out in \cite{us}
the stripe type systems with open boundaries 
show a linear finite size behaviour in $1/N$.
We therefore extrapolate  the $N=16,24,32$ 
stripe systems as a function of $1/N$ and find
$x_c=0.6624$, which is an extrapolation 
for the $4\times n-$ stripe, however it cannot be taken as a
limit for the two-dimensional system, because we deal essentially
with a one-dimensional system.

On first sight it seems a good idea to try an extrapolation of 
the $N=8,18,32$ systems. This is however not feasible,
because the $N=18$ system, has rather pathological periodic 
boundary conditions and cannot be taken as a square system.

The line connecting the system with $N=16$ and $N=36$ is 
not an extrapolation, since the dependence on 
$1/N^{3/2}$, which is motivated by \cite{CHN,HN,Sandvik}
is a guess. The line  crosses the ordinate at
$x_c=0.678$, 
assuming a $1/N$ dependence gives $x_c=0.681$ and 
assuming $1/N^2$ yields $x_c=0.6764$.
All three values lie within the upper and lower bonds
from Ref.\cite{us}.

The fact, that the $x_c$ for $N=16$ is smaller than for $N=36$
prompts us to believe that 
the true $x_c$ should be larger than 
$x_c=0.6744$.
If true, this means, that the value 
$x_c=0.63$ found by Knetter et. al \cite{KnUh} using the flow
equation method is too small. 
We roughly agree with the value 
$x_c=0.7\pm 0.01$ from Ref.\cite{MiUe}, which is obtained by
extrapolating results from systems with $4-20$ spins. 

\subsection{The Plaquette Model }
\label{sec:peri.2}

The finite size behaviour of the plaquette model is in many respects
similar to the Shastry--Sutherland model. 
Again the $4\times n$ stripes have lower $x_c$ than the two-dimensional systems.
Extrapolating the $N=18$ and $N=32$ squares in the above fashion gives 
 $x_c=0.627$ which coincides with the upper bound obtained 
from the $N=16$ system. 
A $1/N$ dependence is not compatible 
with the upper bound and  a $1/N^2$ extrapolation gives $x_c=0.616$.  
It seems more reliable  to draw conclusions from the
$N=36$ and the $N=16$ system.
For the $N=6\times6$ square system we find $x_c=0.59136$,
which is again larger than the corresponding $x_c=0.56198$ 
of the $4\times4$ system. We expect (though again there is no
rigorous argument) that the critical coupling of the infinite size
system should be larger than $x_c=0.59136$.
The line connecting the point of the $4\times4$ and $6\times6$ 
when plotted versus $1/N^{3/2}$ crosses the ordinate at $x_c=0.603$
(assuming a $1/N$ dependence gives $x_c=0.615$ and assuming 
$1/N^2$ yields $x_c=0.599$), all values lie between the rigorous
upper and lower bonds of Sect.\ref{sec:plaq}.

\begin{table}
\begin{center}
{\scriptsize
\begin{tabular}{c|c||c|c}
$J_2/J_1$ & $E_0/N$ & $J_2/J_1$ & $E_0/N$ \\\hline
0.67700000&  -0.374999999998987 & 0.45000000  & 6.99440505513849e-15\\
0.67720000&  -0.374999999999945 & 0.59000000  & 1.49880108324396e-15\\
0.67724000&  -0.374999999997117  & 0.60000000  & -0.00136332822402102\\
0.67724400&  -0.374999999997006  & 0.61000000  & -0.00564902914422299\\
0.67724600&  -0.374999999997179  &  & \\
0.67724800&  -0.375000015810957  &  & \\
0.67730000&  -0.375023759933420 &  & \\
0.67750000& -0.375115180019318  &  & \\
0.67780000& -0.375252606784488  &  & \\
0.68000000& -0.376271688712272  &  & \\
0.70000000& -0.386720856118170  &  & \\
\end{tabular}
}
\end{center}

\caption{Data points calculated for the square $N=32$
 Shastry--Sutherland model (left) and plaquette model (right). The dimension of the subspace containing
  the ground state is  $75.164.451$.}
\label{s32p_2_table}
\end{table}

\begin{table}
\begin{center}
{\scriptsize
\begin{tabular}{c|c||c|c}
$J_2/J_1$ & $E_0/N$ & $J_2/J_1$ & $E_0/N$  \\\hline
0.50000000 &-0.374999999999425 & 0.50000000 & 8.16485767884956e-13\\
0.66697792 &-0.374999999995975 &  0.59135000 & 5.54148393838716e-12\\
0.67441000 &-0.375003966117052 &0.59140000 & -1.70109703724886e-05\\
0.67500000 &-0.375286177720292  & 0.59330700 & -0.000846068815337997\\
0.68000000 &-0.377704740424690  & 0.59331000 & -0.000847382217467496\\
0.70000000 & -0.388578449404367 & 0.62000000 & -0.0133823277140735\\
0.80000000 &-0.451753105004501  & 0.63000000 & -0.0183954992316725\\
0.85000000 &-0.484334145942023  &  & \\
\end{tabular}
} \caption{Data points obtained for the $N=6\times 6=36$
  Shastry--Sutherland model (left) and plaquette model (right). The dimension of
  the subspace with the ground state is $504.174.594$.} \label{s36p_tab}
\end{center}
\end{table}

\section{Summary and Outlook}
\label{sec:summ}

We have introduced a plaquette model with an exactly known ground state
and calculated the critical coupling $x_c$ of the dimer
phase in the model using the
Lanczos algorithm for systems up to $N=36$.
For a $6\times6$ system we find $x_c=0.59136$, 
as rigorous bounds for the critical coupling of the infinite
system we obtain  $0.54263<x_c<0.627(1)$. 

We also calculated the critical coupling of the Shastry--Sutherland model
and obtained $x_c=0.6744$ for a system with periodic 
boundaries and size $N=6\times 6$. 
Extrapolating two-dimensional systems is subtle. Our best guess for the infinite system is $x_c=0.678$ for the Shastry--Sutherland model and
$x_c=0.603$ for the plaquette model.

We hope that our work will prompt further investigations
of the plaquette model, in particular a detailed comparison
of the magnetic properties of the plaquette model
with the "full plaquette interaction" expected in
high $T_c$ mother substances seems an attractive problem.

\begin{table}
\begin{center}
{\scriptsize
\begin{tabular}{c|c|c|c|c|c|}
 system & dimension of subspace & $S_z$ & momentum & reflection
& spin inversion\\\hline
32 square (periodic)&    75.164.451 & $\times$ & $\times$ & & $\times$\\
$4 \times 8$ (periodic)&  37.582.307 & $\times$& $\times$& & $\times$\\
$6 \times 6$ (periodic)& 504.174.594 & $\times$& $\times$& & $\times$\\
28 (open)      &  20.058.300 & $\times$& & & $\times$\\
31 (open)      &   150.283.611 & $\times$& & $\times$&
\end{tabular}
} \caption{Table of the dimensions of the subspaces and symmetry
operations, which were implemented for the 32 square, the $4\times 8$ 
and the $6 \times 6$ system with periodic boundaries and for the 28 and 31 
systems with open boundaries.}
\label{app_tab}
\end{center}
\end{table}

\section{Acknowledgments}
We would like to thank E. M\"uller-Hartmann for many useful discussions.
Most of the data presented were obtained using the 
supercomputing facility IBM p690 (JUMP) at the John von Neumann
Institute for Computing (NIC) at the Forschungszentrum J\"{u}lich 
and the SMP Sun Fire Cluster at the Computing Center of the RWTH Aachen.

\end{document}